# A hybrid lattice Boltzmann and finite difference method for two-phase flows with soluble surfactants

Yan Ba[1], Haihu Liu[2*], Wenqiang Li[1*] and Wenjing Yang[1]

[1]*School of Astronautics, Northwestern Polytechnical University, 127 West Youyi Road, Xi'an 710072, China*

[2]*School of Energy and Power Engineering, Xi'an Jiaotong University, 28 West Xianning Road, Xi'an 710049, China*

\* E-mails: haihu.liu@mail.xjtu.edu.cn (Haihu Liu);

lwq@nwpu.edu.cn (Wenqiang Li)

**Abstract**

A hybrid method is developed to simulate two-phase flows with soluble surfactants. In this method, the interface and bulk surfactant concentration equations of diffuse-interface form, which include source terms to consider surfactant adsorption and desorption dynamics, are solved in the entire fluid domain by the finite difference method, while two-phase flows are solved by a lattice Boltzmann color-gradient model, which can accurately simulate binary fluids with unequal densities. The flow and interface surfactant concentration fields are coupled by a modified Langmuir equation of state, which allows for surfactant concentration beyond critical micelle concentration. The capability and accuracy of the hybrid method are first validated by simulating three numerical examples, including the adsorption of bulk surfactants onto the interface of a stationary droplet, the droplet migration in a constant surfactant gradient, and the deformation of a surfactant-laden droplet in a simple shear flow, in which the numerical results are compared with theoretical solutions and available literature data. Then, the hybrid method is applied to simulate the buoyancy-driven bubble rise in a surfactant solution, in which the influence of surfactants is identified for varying wall confinement, Eotvos number and Biot number. It is found that surfactants exhibit a retardation effect on the bubble rise due to the Marangoni stress that resists interface motion, and the retardation effect weakens as the Eotvos or Biot number increases. We further show that the weakened retardation effect at higher Biot numbers is attributed to a decreased non-uniform effect of surfactants at the interface.

# 1. Introduction

Surfactants are amphiphilic compounds containing a hydrophilic head and a hydrophobic tail. This duality makes them surface active and allows them to alter the interfacial properties of a two-phase system [1]. Surfactants play a critical role in numerous industrial processes, ranging from detergents, food processing, to thermocycling, cell encapsulation and microfluidic flows [2-4]. In particular, surfactants are often applied to manipulate the droplet behavior in microfluidics, and under the influence of surfactants, droplets with smaller size, higher formation frequency and better stability tend to be generated in microfluidic devices [5, 6]. Surfactants can be typically classified into two categories: insoluble and soluble surfactants. Insoluble surfactants exist only at the interface, while soluble surfactants are present both at the interface and in the bulk fluid, which are normally added to at least one bulk fluid, and then migrate to the interface by the adsorption process. This paper aims to develop a numerical method for interfacial flows with soluble surfactants and investigate the role of soluble surfactants on the droplet deformation and breakup, hopefully providing beneficial guidance for functionalizing surfactants and optimizing the choice of surfactants in microfluidics.

Numerical modelling of interfacial flows with soluble surfactants, which includes the coupling of bulk surfactants, interface surfactants and two-phase flow field, is a challenging task. The flow field influences the surfactant distributions by the advection process, and the interface surfactant field in turn affects the interfacial behavior by the interfacial tension reduction and non-uniform effects including non-uniform interfacial tension and Marangoni stresses [7]. In addition, the non-linear coupling system has to be solved over a moving, deforming and even typologically changing interface, and the adsorption of surfactants to, and the desorption from the interface must be dynamically considered. Therefore, in order to simulate interfacial flows with soluble surfactants, the surfactant transport equations should be solved for both interface and bulk phase, and the bulk surfactant field has to be coupled with the interface surfactant field through the adsorption and desorption related source terms.

Modelling and simulation of two-phase flows with soluble surfactants can be tracked back to work of Stone [8], who proposed a convection-diffusion equation for the transport of insoluble surfactants in the framework of sharp interface. Based on his pioneering work [8], a number of numerical methods have been developed to simulate two-phase flows with insoluble surfactants, which include the boundary integral method (BIM) [9, 10], front-tracking method [11], level-set method [12], and the volume-of-fluid (VOF) method [13]. With great success in simulating insoluble surfactants, numerical methods based on the sharp-interface description were also established to simulate droplet dynamics with soluble surfactants. Milliken and Leal [14] conducted a pioneering work to investigate the effect of surfactant solubility on the deformation of a viscous droplet in a uniaxial extensional flow using the BIM, which includes the transport mechanisms of soluble surfactants like the bulk diffusion, interface diffusion and convection, but ignores the bulk convection. Muradoglu and Tryggvason [15] proposed a front-tracking method for the simulation of two-phase flows with soluble surfactants, in which the adsorption and desorption dynamics were modeled in a conservative way by distributing the adsorbed surfactants as a source term in the bulk surfactant concentration equation. Such a treatment was later adopted extensively in the front-tracking method [16, 17], arbitrary Lagrangian-Eulerian finite-element method [18], and the level-set method [19]. Chen and Lai [20] proposed a conservative scheme for solving coupled interface-bulk surfactant concentration equations based on the immersed boundary method. In their method, an indicator function was used to map the original bulk surfactant concentration equation in an irregular domain into a regular computational domain. Khatri and Tornberg [21] presented a two-dimensional (2D) method for two-phase flows with soluble surfactants, which uses an embedded boundary method to deal with the boundary condition for soluble surfactants at the dynamically deformed interface, a segment projection method to represent the interface, and a finite difference method on the regular grids for the bulk surfactants. Hu and Lai [22] proposed a coupled immersed interface and immersed boundary method for two-phase flows with soluble surfactants, in which the immersed interface method is used to solve the interface and bulk surfactant equations.

Although numerical methods for simulating interfacial flows with soluble surfactants have gained great success, they still suffer from some shortcomings. For example, the BIM and front-tracking method often encounter the difficulties of handling surface mesh distortion and topological changes such as interface rupture and coalescence. The VOF and level-set methods require interface reconstruction algorithms or re-initialization process to represent or correct the interface, which may be complex or lead to unphysical outcomes. In addition, in most of these methods, the interface surfactant concentration has to be extended artificially to the neighborhood of the interface to evaluate the derivatives in the surfactant concentration equations [12].

In addition to the sharp-interface methods, the diffuse-interface methods have also been a popular tool to simulate two-phase flows with soluble surfactants [23-34]. Among various diffuse-interface methods, the phase-field (PF) method proposed by Teigen et al. [24] is the most representative. In their method, the surfactant dynamics is modelled by solving the surfactant evolution equations of diffuse-interface form, which are obtained by extending the interface and bulk surfactant concentration equations of sharp-interface form to the entire fluid domain with the aid of a delta function. The PF method, however, is often criticized for unphysical dissolution of small droplets and the dependence of numerical results on mobility [35, 36]. Xu et al. [37] proposed a level-set method for interfacial flows with soluble surfactant, where the bulk surfactant concentration equation defined in irregular moving domain was solved by the use of the diffusive domain method. As a level-set method, it also inherits the drawback of not being able to conserve the mass for each fluid. Recently, a hybrid lattice-Boltzmann and finite-difference (LB-FD) method was developed by Liu et al. [7] to simulate two-phase flows with insoluble surfactants, which uses the LB color-gradient model to solve two-phase flows and the finite difference method to solve the surfactant transport, described by the interface surfactant concentration equation of diffuse-interface form [24]. The LB-FD method inherits the advantages of the LB color-gradient model, such as the ease to implement, the flexibility in adjusting interfacial tension, strict mass conservation for each fluid, and the ability of dealing with topological changes of interface [38-42], while allowing one to solve the transport of insoluble surfactants without additional extension or initialization procedures. In addition, it was later extended to the modelling of contact-line dynamics with insoluble surfactants [43, 44].

In the present work, we will extend the hybrid LB-FD method [7] to the simulation of droplet interfacial dynamics with soluble surfactants. Unlike Liu et al. [7], the present method allows for two-phase fluids with unequal densities, which is achieved by an improved color-gradient model proposed by Ba et al. [42]. To account for the surfactant solubility, an additional convection-diffusion equation is introduced to describe the transport of bulk surfactants, which, like the interface concentration equation, is also expressed in the diffuse-interface form. A source term related to the adsorption and desorption dynamics is added to both interface and bulk surfactant concentration equations to consider the transfer of surfactants across the interface. The capability and accuracy of the hybrid method for soluble surfactants are first assessed by three numerical examples, namely the adsorption of bulk surfactants onto the interface of a three-dimensional (3D) stationary droplet, 3D droplet migration in a surfactant concentration gradient, the deformation of a surfactant-covered droplet under 2D shear flow. It is then used to investigate the influence of soluble surfactants on the 3D bubble rise.

## 2 Numerical Method

A hybrid LB-FD method is developed for simulating the immiscible two-phase flow with soluble surfactants, which is built upon the recent work by Liu et al. [7]. To exceed the limit of equal densities, the method uses an improved LB color-gradient model [42, 45] to simulate two-phase flows, in which the Marangoni force caused by uneven surfactant distribution is additionally incorporated through the perturbation operator [7]. Considering the diffuse nature of the interface obtained from the color-gradient model, the interface and bulk convection-diffusion equations proposed by Teigen et al. [24]

are introduced to describe the transport of soluble surfactants, which are solved in the entire fluid domain by a finite difference method. The LB model and finite difference method are coupled by a modified Langmuir equation of state [44, 46], which relates interfacial tension to interface surfactant concentration and allows for the surfactant concentration exceeding the critical micelle concentration (CMC).

*2.1 An improved LB color-gradient model for immiscible two-phase flows*

In the color-gradient model, we consider two immiscible fluids (say red and blue fluids), and use the distribution function $f_i^k$ to represent the fluid $k$, where $k = R$ or $B$ denotes the color "red" or "blue", and $i$ is the lattice velocity direction ranging from 0 to ($n$-1) for a D$m$Q$n$ lattice model. In this study, the D2Q9 and D3Q19 lattice models are used for 2D and 3D simulations, respectively.

In each time step, the total distribution function, defined as $f_i = f_i^R + f_i^B$, first experiences a collision step as

$$f_i^\dagger(\mathbf{x},t) = f_i(\mathbf{x},t) + \Omega_i(\mathbf{x},t) + \Phi_i, \tag{1}$$

where $f_i(\mathbf{x},t)$ is the total distribution function at the position $\mathbf{x}$ and time $t$ in the $i$-th lattice velocity direction, $\Omega_i$ is the collision operator, $\Phi_i$ is the perturbation operator responsible for creating normal capillary force and tangential Marangoni stress, and $f_i^\dagger$ is the post-collision distribution function.

To recover the correct Navier-Stokes equations for two-phase flows with unequal densities, we use a multiple-relaxation-time (MRT) collision operator with the source term, previously proposed by Ba et al. [42] in 2D and Wen et al. [45] in 3D. The MRT collision operator reads as

$$\Omega_i = \sum_k \sum_j \left[ -\left(\mathbf{M}^{-1}\mathbf{S}\right)_{ij} \left(m_j^k - m_j^{k,eq}\right) + \left(\mathbf{M}^{-1} - \frac{1}{2}\mathbf{M}^{-1}\mathbf{S}\right)_{ij} C_j^k \delta_t \right], \tag{2}$$

where $\mathbf{S}$ is a diagonal relaxation matrix. $\mathbf{M}$ is a linear transformation matrix projecting the distribution function $\mathbf{f}^k$ in the discrete velocity space onto $\mathbf{m}^k$ in the moment space, i.e. $\mathbf{m}^k = \mathbf{M}\mathbf{f}^k$, where $\mathbf{f}^k = \left(f_0^k, f_1^k, \cdots, f_{n-1}^k\right)^T$ and $\mathbf{m}^k = \left(m_0^k, m_1^k, \cdots, m_{n-1}^k\right)^T$. $\mathbf{m}^{k,eq} = \left(m_0^{k,eq}, m_1^{k,eq}, \cdots, m_{n-1}^{k,eq}\right)^T$ is the equilibria in the moment space, and it is related to the equilibrium distribution function $\mathbf{f}^{k,eq}$ by $\mathbf{m}^{k,eq} = \mathbf{M}\mathbf{f}^{k,eq}$, where $\mathbf{f}^{k,eq}$ can be constructed by the third-order Hermite expansion of the Maxwellian distribution [47, 48]. $\mathbf{C}^k$ is the correction term in the moment space, and $C_j^k$ is its $j$-th element. For the D2Q9 and D3Q19 models, the expressions of $\mathbf{M}$, $\mathbf{S}$, $\mathbf{m}^{k,eq}$ and $\mathbf{C}^k$ can be found in Refs. [42, 45], which are also presented in Appendix A for the convenience of the reader.

The perturbation operator $\Phi_i$ in Eq.(1) is also implemented in the moment space using a forcing term as [42, 45]

$$\Phi_i = \sum_j \left( \mathbf{M}^{-1} - \frac{1}{2}\mathbf{M}^{-1}\mathbf{S} \right)_{ij} F_j \tag{3}$$

with

$$F_j = \sum_k \mathbf{M}_{jk} w_k \left[ \frac{\mathbf{e}_k - \mathbf{u}}{c_s^2} + \frac{(\mathbf{e}_k \cdot \mathbf{u})\mathbf{e}_k}{c_s^4} \right] \cdot \mathbf{F}_s(\mathbf{x},t), \tag{4}$$

where $\mathbf{u}$ is the local fluid velocity, and $\mathbf{F}_s$ is the interfacial force consisting of the capillary force and the Marangoni stress. For the D2Q9 and D3Q19 models used, the speed of sound $c_s$ is $1/\sqrt{3}$, and the weight factors $w_i$ and the discrete velocity vectors $\mathbf{e}_i$ can be found in the previous works [42, 45]. Based on the continuum surface force model [49, 50], the interfacial force $\mathbf{F}_s$ is given by [51]

$$\mathbf{F}_s(\mathbf{x},t) = \frac{1}{2} |\nabla \rho^N| \left( \sigma \kappa \mathbf{n} + \nabla_s \sigma \right) \tag{5}$$

with the color function $\rho^N$ distinguishing two different fluids defined as [42]

$$\rho^N = \left( \frac{\rho^R}{\rho^{R0}} - \frac{\rho^B}{\rho^{B0}} \right) \Big/ \left( \frac{\rho^R}{\rho^{R0}} + \frac{\rho^B}{\rho^{B0}} \right), \quad -1 \leq \rho^N \leq 1, \tag{6}$$

where $\rho^{R0}$ and $\rho^{B0}$ are the densities of the pure red and blue fluids, and $\rho^R$ and $\rho^R$ are the local densities of the red and blue fluids, respectively. $\nabla_s = (\mathbf{I} - \mathbf{nn}) \cdot \nabla$ is the surface gradient operator, and $\mathbf{I}$ is the second-order identity tensor. $\sigma$ is the interfacial tension coefficient, $\mathbf{n} = -\nabla \rho^N / |\nabla \rho^N|$ is the unit vector normal to the interface, and $\kappa$ is the local interface curvature related to $\mathbf{n}$ by

$$\kappa = -\nabla_s \cdot \mathbf{n} = -\nabla \cdot \mathbf{n}. \tag{7}$$

In the presence of soluble surfactants, an equation of state (EOS) is needed to relate the interfacial tension coefficient $\sigma$ to the interface surfactant concentration $\psi$. Here, a modified Langmuir EOS [44, 46, 52] is used, and it reads as

$$\sigma(\psi) = \max\{\sigma_0[1 + E_0 \ln(1 - \psi/\psi_\infty)], \sigma_{min}\}, \tag{8}$$

where $\psi_\infty$ is the maximum capacity of the interface surfactant concentration, $E_0 = RT\psi_\infty/\sigma_0$ is the elasticity number with $\sigma_0$ being the interfacial tension coefficient at $\psi = 0$ (i.e., clean interface), $R$ is the ideal gas constant, and $T$ is the absolute temperature. In this study, $\sigma_{min}$ is set to $\sigma_0/10$, and thus one can compute the CMC, i.e., $CMC = \psi_\infty[1 - \exp(-0.9/E_0)]$. Eq.(10) suggests that the interfacial tension would remain a constant of $\sigma_{min}$ when the interface surfactant concentration exceeds the CMC, consistent with the previous studies [44, 46]. As previously done in Refs. [7, 12, 24], the initial interface surfactant concentration $\psi_0$ is quantified by the surfactant coverage $x_{in}$, i.e. $x_{in} = \psi_0/\psi_\infty$.

Substituting Eq. (8) into Eq. (5), the interfacial force can be expressed as

$$\mathbf{F}_s = \begin{cases} -\dfrac{1}{2}\sigma\kappa\nabla\rho^N - \dfrac{1}{2}|\nabla\rho^N|\dfrac{\sigma_0 E_0}{\psi_\infty - \psi}[\nabla\psi - (\mathbf{n}\cdot\nabla\psi)\mathbf{n}], & \psi < CMC \\ -\dfrac{1}{2}\sigma_{min}\kappa\nabla\rho^N, & \psi \geq CMC \end{cases} \tag{9}$$

where all the partial derivatives including those in the curvature calculation are evaluated by the fourth-order isotropic finite difference scheme [40].

With the perturbation operator given by Eqs.(5) and (6), the local fluid velocity should be defined as [40, 42]

$$\rho\mathbf{u}(\mathbf{x},t) = \sum_i f_i(\mathbf{x},t)\mathbf{e}_i + \frac{1}{2}\mathbf{F}_s(\mathbf{x},t) \tag{10}$$

After the collision step, a recoloring step is performed to promote the phase segregation and maintain a reasonable interface. Following Latva-Kokko and Rothman [53], the recoloring step reads as

$$f_i^{R\ddagger}(\mathbf{x},t) = \frac{\rho^R}{\rho}f_i^\dagger(\mathbf{x},t) + \beta w_i \frac{\rho^R \rho^B}{\rho}\cos(\theta_i)|\mathbf{e}_i|;$$
$$f_i^{B\ddagger}(\mathbf{x},t) = \frac{\rho^B}{\rho}f_i^\dagger(\mathbf{x},t) - \beta w_i \frac{\rho^R \rho^B}{\rho}\cos(\theta_i)|\mathbf{e}_i|, \tag{11}$$

where $f_i^{k\ddagger}$ is the recolored distribution function of the fluid k, and $\theta_i$ is the angle between $\nabla\rho^N$ and $\mathbf{e}_i$ defined as

$$\cos(\theta_i) = \frac{\mathbf{e}_i \cdot \nabla\rho^N}{|\mathbf{e}_i||\nabla\rho^N|}. \tag{12}$$

$\beta$ is the segregation parameter, which is fixed at 0.7 to resolve the interface and at the same time keep the interface sharp [40, 50, 54].

Finally, the red and blue distribution functions both undergo a propagation step as

$$f_i^k(\mathbf{x}+\mathbf{e}_i\delta_t, t+\delta_t) = f_i^{k\ddagger}(\mathbf{x},t), \quad k = R, B, \tag{13}$$

and the resulting distribution functions are applied to calculate the fluid densities through $\rho^k = \sum_i f_i^k$.

## *2.2 Finite difference method for the transport of soluble surfactants*

With a sharp interface representation, Milliken and Leal [14] investigated the effect of surfactant solubility on the droplet deformation and breakup using the BIM, in which the interface surfactant concentration is governed by a convection-diffusion equation along the interface, as proposed by Stone [8], and the bulk surfactant concentration is assumed to be governed by diffusion only:

$$\partial_t \psi + \nabla_s \cdot (\psi \mathbf{u}_s) + \kappa \psi \mathbf{u} \cdot \mathbf{n} = D_s \nabla_s^2 \psi + j_s, \tag{14}$$

$$\nabla^2 \phi = 0, \tag{15}$$

where $\mathbf{u}_s = (\mathbf{I} - \mathbf{nn}) \cdot \mathbf{u}$ is the fluid velocity tangential to the interface, and $D_s$ is the interface surfactant diffusivity. The source term in Eq.(14) is defined as $j_s = r_a \phi(\mathbf{x}_s) - r_d \psi(\mathbf{x}_s)$, in which $r_a$ and $r_d$ are the adsorption and desorption coefficients, respectively. $\mathbf{x}_s$ is the position adjacent to the interface which is often referred to as the "sublayer", and $\phi$ is the bulk surfactant concentration. The boundary condition imposed at the interface for the bulk concentration equation can be obtained by the mass conservation, and it reads as $\nabla \phi \cdot \mathbf{n} = r_a r / D_b$, where $r$ is the droplet radius and $D_b$ is the bulk surfactant diffusivity.

Using a general form of the bulk surfactant concentration equation, namely

$$\partial_t \phi + \nabla \cdot (\phi \mathbf{u}) = D_b \nabla \cdot (\nabla \phi) + j_b, \tag{16}$$

Muradoglu and Tryggvason [15] presented a front-tracking method for two-phase flows with soluble surfactants, in which the interface surfactant concentration equation is the same as the one used by Milliken and Leal [14] but with a new definition of the interface source term, given by

$$j_s = r_a \phi_s (\psi_\infty - \psi) - r_d \psi. \tag{17}$$

In the above equations, $\phi_s$ is the bulk surfactant concentration adjacent to the interface, and $j_b$ is the bulk source term at the interface which is distributed onto the adsorption layer near the interface in a conservative way. With the aid of $j_b$, the boundary condition at the interface can be simplified as

$$\mathbf{n} \cdot \nabla \phi = 0.$$

By extending Eqs. (14) and (16) to the entire fluid domain, Teigen et al. [24] proposed the surfactant concentration equations of the diffuse-interface form, which are expressed as

$$\partial_t(\delta_\Gamma \psi) + \nabla \cdot (\delta_\Gamma \psi \mathbf{u}) = D_s \nabla \cdot (\delta_\Gamma \nabla \psi) + \delta_\Gamma j, \tag{18}$$

$$\partial_t(\chi \phi) + \nabla \cdot (\chi \phi \mathbf{u}) = D_b \nabla \cdot (\chi \nabla \phi) - \delta_\Gamma j, \tag{19}$$

for the interface and bulk surfactant concentrations, respectively. Herein, $\delta_\Gamma$ is the Dirac function which satisfies $\int_\Gamma \psi d\Gamma = \int_\Omega \psi \delta_\Gamma d\Omega$. $\chi$ is to identify the bulk phase with soluble surfactants, and it is assumed to have $\chi = 1$ in the bulk phase with soluble surfactants and $\chi = 0$ elsewhere [24]. $j$ is the source term denoting the net flux of surfactants and has the same definition as $j_s$ in Eq.(17). Eqs.(18) and (19) are both solved over the whole computational domain, which allows for the evolution of surfactants without any special treatments [7, 24]. By coupling the surfactant transport, i.e. Eqs.(18) and (19), with the phase-field method, Teigen et al. [24] successfully simulated two-phase flows with soluble surfactants.

Similar to the phase-field method, the LB color-gradient model is also a diffuse-interface model, so Eqs. (18) and (19) are adopted to describe the surfactant transport in the present method. Following the previous work [7], we take $\delta_\Gamma = |\nabla \rho^N|/2$ and choose $\chi = (1 - \rho^N)/2$ so that the bulk surfactants are present in the blue fluid only. Thus, the surfactant evolution equations are expressed as

$$\partial_t(|\nabla \rho^N| \psi) + \nabla \cdot (|\nabla \rho^N| \psi \mathbf{u}) = D_s \nabla \cdot (|\nabla \rho^N| \nabla \psi) + |\nabla \rho^N| j, \tag{20}$$

$$\partial_t[(1 - \rho^N)\phi] + \nabla \cdot [(1 - \rho^N)\phi \mathbf{u}] = D_b \nabla \cdot [(1 - \rho^N) \nabla \phi] - |\nabla \rho^N| j \tag{21}$$

with
$$j = r_a \phi(\psi_\infty - \psi) - r_d \psi, \tag{22}$$

where we have replaced $\phi_s$ in Eq.(17) with $\phi$. Such a treatment avoids the need to search the nodes adjacent to the interface for obtaining the corresponding concentration values, thus simplifying the simulation process.

The interface and bulk surfactant concentration equations are solved by the finite difference method, in which a modified Crank-Nicolson scheme [55] is first applied for the time discretization. The equations after the time discretization read as

$$\frac{|\nabla \rho^N|^{t+\delta_t} \psi^{t+\delta_t} - |\nabla \rho^N|^t \psi^t}{\delta_t} = D_s \frac{\nabla \cdot (|\nabla \rho^N| \nabla \psi)^{t+\delta_t} + \nabla \cdot (|\nabla \rho^N| \nabla \psi)^t}{2} + S^t,$$

$$\frac{[(1-\rho^N)\phi]^{t+\delta_t} - [(1-\rho^N)\phi]^t}{\Delta t} = D_b \frac{\nabla \cdot [(1-\rho^N)\nabla \psi]^{t+\delta_t} + \nabla \cdot [(1-\rho^N)\nabla \psi]^t}{2} + B^t, \tag{23}$$

for $t = 0$ and

$$\frac{|\nabla \rho^N|^{t+\delta_t} \psi^{t+\delta_t} - |\nabla \rho^N|^t \psi^t}{\delta_t} = D_s \frac{\nabla \cdot (|\nabla \rho^N| \nabla \psi)^{t+\delta_t} + \nabla \cdot (|\nabla \rho^N| \nabla \psi)^t}{2} + \frac{3}{2} S^t - \frac{1}{2} S^{t-\delta_t},$$

$$\frac{[(1-\rho^N)\phi]^{t+\delta_t} - [(1-\rho^N)\phi]^t}{\delta_t} = D_b \frac{\nabla \cdot [(1-\rho^N)\nabla \phi]^{t+\delta_t} + \nabla \cdot [(1-\rho^N)\nabla \phi]^t}{2} + \frac{3}{2} B^t - \frac{1}{2} B^{t-\delta_t},$$
(24)

for $t \geq 1$, where $S = -\nabla \cdot (\psi |\nabla \rho^N| \mathbf{u}) + |\nabla \rho^N| j$ and $B = -\nabla \cdot [\phi(1-\rho^N)\mathbf{u}] - |\nabla \rho^N| j$. The spatial discretization is further needed to solve Eqs.(26) and (27). Specifically, for the convection terms $\nabla \cdot (\psi |\nabla \rho^N| \mathbf{u})$ and $\nabla \cdot [\phi(1-\rho^N)\mathbf{u}]$, a third-order weighted essentially nonoscillatory scheme is used. Taking $\nabla \cdot (\psi |\nabla \rho^N| \mathbf{u})$ as an example, and denoting $\mathbf{u} = (u_x, u_y, u_z)$ and $\mathbf{\Psi} = (\Psi_x, \Psi_y, \Psi_z) = \psi |\nabla \rho^N| \mathbf{u}$, we can evaluate $\nabla \cdot \mathbf{\Psi}$ at the grid point $(x_i, y_j, z_k)$ as

$$\nabla \cdot \mathbf{\Psi}|_{ijk} = \frac{u_x^+}{|u_x|} D_x^- \Psi_{x,ijk} + \frac{u_x^-}{|u_x|} D_x^+ \Psi_{x,ijk} + \frac{u_y^+}{|u_y|} D_y^- \Psi_{y,ijk} + \frac{u_y^-}{|u_y|} D_y^+ \Psi_{y,ijk} + \frac{u_z^+}{|u_z|} D_z^- \Psi_{z,ijk} + \frac{u_z^-}{|u_z|} D_z^+ \Psi_{z,ijk}, \quad (25)$$

where $x^+ = \max(x, 0)$, $x^- = \min(x, 0)$, and $D_x^\pm \Psi_{x,ijk}$, $D_y^\pm \Psi_{y,ijk}$ and $D_z^\pm \Psi_{z,ijk}$ are the one-sided divided differences in the x, y and z directions, respectively [55]. For example, the approximation to $D_x^- \Psi_{x,ijk}$ on the left-biased stencil $\{x_{i-2}, x_{i-1}, x_i, x_{i+1}\}$ is

$$D_x^- \Psi_{x,ijk} = \frac{1}{2\delta_x} \left[ (\Delta^+ \Psi_{x,(i-1)jk} + \Delta^+ \Psi_{x,ijk}) - w_- (\Delta^+ \Psi_{x,(i-2)jk} - 2\Delta^+ \Psi_{x,(i-1)jk} + \Delta^+ \Psi_{x,ijk}) \right] \quad (26)$$

with $w_- = 1/(1+2r_-^2)$ and $r_- = \frac{\epsilon + (\Delta^+\Delta^- \Psi_{x,(i-1)jk})^2}{\epsilon + (\Delta^+\Delta^- \Psi_{x,ijk})^2}$, where $\Delta^+$ and $\Delta^-$ are the forward and backward difference operators, respectively; $\epsilon$ is a small positive number and is set to $10^{-6}$ in our computations [55]. By the symmetry, the approximation to $D_x^+ \Psi_{x,ijk}$ on the right-biased stencil $\{x_{i-1}, x_i, x_{i+1}, x_{i+2}\}$ is

$$D_x^+ \Psi_{x,ijk} = \frac{1}{2\delta_x} \left[ (\Delta^+ \Psi_{x,(i-1)jk} + \Delta^+ \Psi_{x,ijk}) - w_+ (\Delta^+ \Psi_{x,(i+1)jk} - 2\Delta^+ \Psi_{x,ijk} + \Delta^+ \Psi_{x,(i-1)jk}) \right], \quad (27)$$

where $w_+ = 1/(1+2r_+^2)$ with $r_+ = \frac{\epsilon + (\Delta^+\Delta^- \Psi_{x,(i+1)jk})^2}{\epsilon + (\Delta^+\Delta^- \Psi_{x,ijk})^2}$. The diffusion term $\nabla \cdot (|\nabla \rho^N| \nabla \psi)$ can be discretized as

$$\nabla \cdot (|\nabla \rho^N| \nabla \psi)_{i,j,k} = |\nabla \rho^N|_{i-\frac{1}{2},j,k} \psi_{i-1,j,k} + |\nabla \rho^N|_{i+\frac{1}{2},j,k} \psi_{i+1,j,k}$$
$$+ |\nabla \rho^N|_{i,j-\frac{1}{2},k} \psi_{i,j-1,k} + |\nabla \rho^N|_{i,j+\frac{1}{2},k} \psi_{i,j+1,k} + |\nabla \rho^N|_{i,j,k-\frac{1}{2}} \psi_{i,j,k-1} + |\nabla \rho^N|_{i,j,k+\frac{1}{2}} \psi_{i,j,k+1} \quad (28)$$
$$- \left( |\nabla \rho^N|_{i-\frac{1}{2},j,k} + |\nabla \rho^N|_{i+\frac{1}{2},j,k} + |\nabla \rho^N|_{i,j-\frac{1}{2},k} + |\nabla \rho^N|_{i,j+\frac{1}{2},k} + |\nabla \rho^N|_{i,j,k-\frac{1}{2}} + |\nabla \rho^N|_{i,j,k+\frac{1}{2}} \right) \psi_{i,j,k},$$

where $|\nabla \rho^N|_{i-\frac{1}{2},j,k} = \frac{1}{2} \left( |\nabla \rho^N|_{i,j,k} + |\nabla \rho^N|_{i-1,j,k} \right)$, and the other terms containing $1/2$ points are also obtained by interpolation. A similar treatment can be used for the discretization of the diffusion term $\nabla \cdot [(1-\rho^N)\nabla \phi]$.

The resulting linear systems for $\psi^{t+\delta_t}$ and $\phi^{t+\delta_t}$ are symmetric positive definite, and can be solved by the successive over-relaxation method, in which the relaxation factor is taken as 1.2. To conserve the total mass of surfactants, the rescaling technique proposed by Xu et al. [37] is used, which is simple and effective.

## 3 Numerical Results and Validations

In this section, three numerical examples, namely the adsorption dynamics of bulk surfactants onto a droplet interface, the droplet migration driven by a constant surfactant gradient, and the deformation of a 2D surfactant-covered droplet under simple shear flow, are first conducted to assess the ability of the present method in dealing with interfacial flows with soluble surfactants, in which the simulation results are quantitatively compared with theoretical solutions and available literature data. Then, the hybrid method is applied to investigate 3D bubble rise in a surfactant solution, in which the influence of surfactants is identified for varying wall confinement, Eotvos number and Biot number.

### *3.1 Adsorption dynamics of bulk surfactants*

Here, we consider a spherical droplet (red fluid) with radius $R$ in a static cubic domain filled with a second fluid (blue fluid), and study the adsorption dynamics of bulk surfactants in the blue fluid onto the droplet interface. Due to the symmetry of the problem, one octant of the droplet is simulated with the domain size of $5R \times 5R \times 5R$. Symmetric boundary conditions [7] are applied at the bottom, front and left boundaries, and periodical boundary conditions are used at the other boundaries. The droplet is centered at the origin of the coordinates, i.e., $(x,y,z) = (0,0,0)$. Initially, the droplet interface is assumed to be clean, i.e., $\psi_0 = 0$, and bulk surfactants are evenly distributed in the blue fluid with $\phi_0 = 0.0005$. The fluid properties are set as $\rho^{R0} = \rho^{B0} = 1$, $\mu^{R0} = \mu^{B0} = 0.1$ and $\sigma = 0.1$, where $\mu^{R0}$ and $\mu^{B0}$ are the dynamic viscosities of the pure red and blue fluids, respectively, and they appear in Eq.(43). The interface and bulk surfactant diffusivities are $D_s = 0.1$ and $D_b = 0.1$. The elasticity number is $E_0 = 0.5$. A simple source term $j = r_a \psi_\infty \phi$ with $r_a = 0.2$ is applied so that the surfactants are only transported from the bulk to the interface.

As indicated by Muradoglu and Tryggvason [15], in an infinite domain or for a short period, the dimensionless bulk surfactant concentration $\phi/\phi_0$ and the dimensionless interface surfactant concentration $\psi/\psi_\infty$ analytically follow

$$\frac{\phi}{\phi_0} = 1 - \frac{r_a \psi_\infty \sqrt{\pi D_s t}/D_s}{1+\sqrt{\pi D_s t}(1+r_a \psi_\infty R/D_s)} \frac{R}{r} erfc\left(\frac{r-R}{2\sqrt{D_s t}}\right), \tag{29}$$

and

$$\psi/\psi_\infty = \psi_0/\psi_\infty + r_a \phi_0 \left[ t - \frac{am}{\eta^3}\left(\eta^2 t - 2\eta\sqrt{t} + 2\ln(1+\eta\sqrt{t})\right)\right] \tag{30}$$

where $erfc(x)$ is the complementary error function and $r$ is local distance to the droplet center; $m = r_a \psi_\infty / D_s$, $a = \sqrt{\pi D_s}$ and $\eta = a(1+Rm)/R$.

We first conduct the grid independence test with three different grid resolutions, i.e., $R = 20$, $R = 30$ and $R = 40$. Figure 1 presents the time evolution of $\psi/\psi_\infty$ at $\rho^N = 0$ for three different grid resolutions, where the analytical results are also shown for comparison. It is seen that as the dimensionless time $\tau = D_b t/a^2$ increases, $\psi/\psi_\infty$ increases in a nearly linear manner, especially when $\tau \geq 1$. Also, as the grid resolution changes from $R = 30$ to 40, invisible difference between the results is noticed. This indicates that the grid resolution with $R = 30$ is fine enough to produce grid-independent results, and thus it will be used in the subsequent simulations.

Figure 2 shows the snapshots of the dimensionless bulk surfactant concentration, defined as $\phi^* = \phi/\phi_0$, at $y = 0$ plane for $\tau = 0.11$, 0.56, 1.56 and 3. It is seen that at the early stage of the simulation ($\tau = 0.11$), the bulk surfactant concentration $\phi^*$ is unity in the blue fluid, and changes from 0 to 1 adjacent to the interfacial region. As time increases ($\tau = 0.56$ and 1.56), more surfactants are adsorbed onto the interface and the bulk surfactant concentration neighboring to the interface gradually decreases. We also notice at $\tau = 3$ that the region of low bulk surfactant concentration spreads continuously since the desorption is not considered. In Figure 3, $\phi^*$ along the radial direction at different time is illustrated, and for the purpose of comparison, the analytical results from Eq.(29) are also plotted. It is clear that the profiles of the bulk surfactant concentration match perfectly with the analytical ones for $\tau \leq 1.11$, but when reaching $\tau = 3$, the bulk surfactant concentrations obtained from the present method are lower than the analytical ones for $r/R > 3.5$, which is because the domain size is not large enough or the simulation time is too long to satisfy the infinite domain or short period assumption made in the derivation of the analytical solutions.

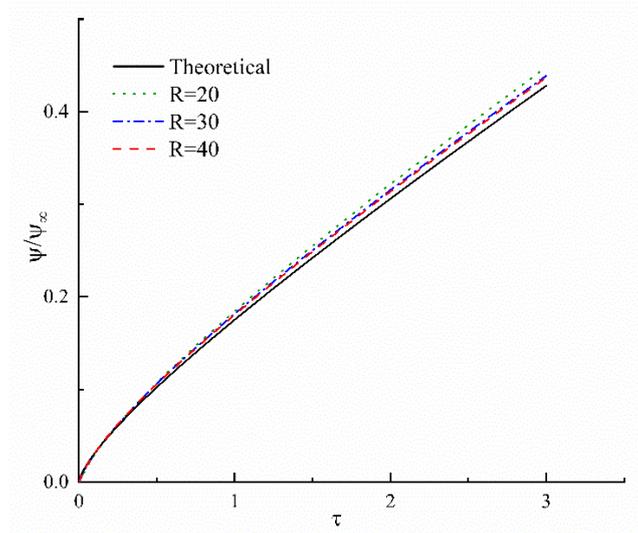

Figure 1 Time evolution of the dimensionless interface surfactant concentration $\psi/\psi_\infty$ at $\rho^N = 0$ for the grid resolutions of $R = 20$, 30 and 40. The theoretical results, represented by the solid lines, are also plotted for comparison. The dimensionless time is defined as $\tau = D_b t / a^2$.

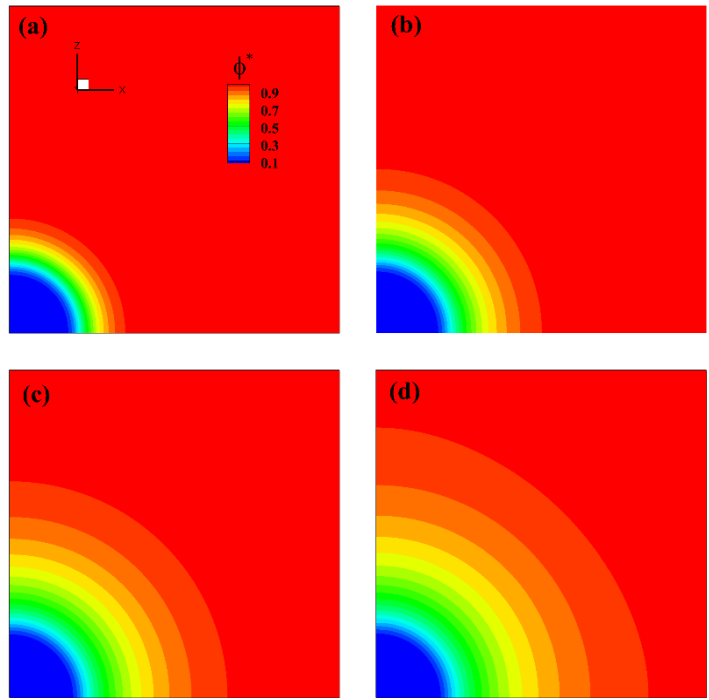

Figure 2 The distributions of the dimensionless bulk surfactant concentration $\phi^*$ at $y = 0$ plane for $\tau$ of (a) 0.11, (b) 0.56, (c) 1.56 and (d) 3. The black lines in each subfigure represent the interface

$$\rho^N = 0.$$

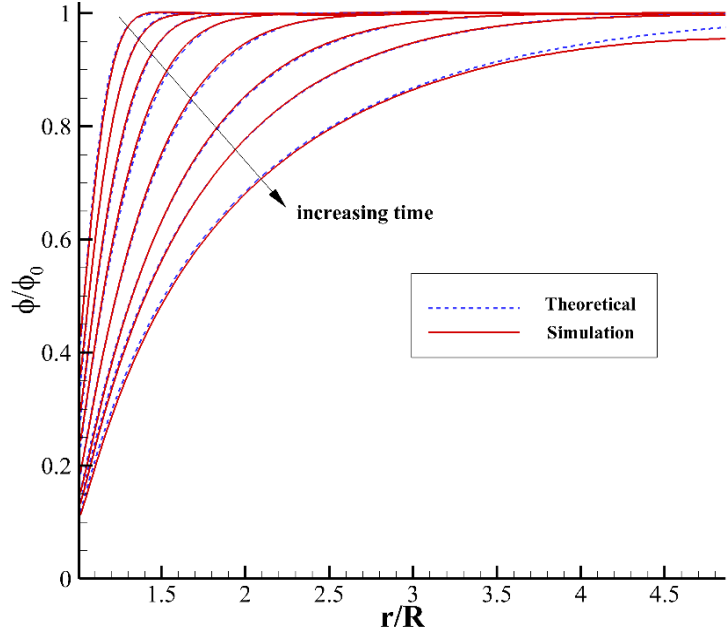

Figure 3 The distributions of $\phi^* = \phi/\phi_0$ along the radial direction at $\tau = 0.011$, 0.022, 0.044, 0.089, 0.22, 0.55, 1.11, and 3. The red solid lines represent the simulation results, while the dashed lines are the theoretical solutions given by Eq.(32).

### *3.2 Droplet migration driven by a constant surfactant gradient*

In this section, we perform a test for the droplet migration driven by a constant surfactant gradient. Instead of imposing a temperature gradient as done in Young et al. [51, 56], we vary the interface surfactant concentration in the vertical (*z*) direction that leads to the change in the interfacial tension and thus generates the Marangoni stresses. With a linear distribution of the interface surfactant concentration, namely

$$\psi(z) = \psi_\infty z / L_z, \tag{31}$$

and assuming $\sigma(\psi) = \sigma_0(1 - E_0 \psi / \psi_\infty)$, the droplet (red fluid) will migrate upward and eventually reach a terminal velocity. In an unbounded blue fluid domain, Young et al. [16, 56] derived an analytical solution for the terminal velocity, also known as the YGB velocity, which reads as:

$$U_{YGB} = \frac{2\sigma_0 E_0 R}{L_z \left(6\mu^{R0} + 9\mu^{B0}\right)}, \tag{32}$$

where $L_z$ is the size of the computational domain in the $z$ direction, and $R$ is the droplet radius. The simulations are conducted with two different sizes of computational domain, i.e. $L_x \times L_y \times L_z = 5R \times 5R \times 10R$ and $L_x \times L_y \times L_z = 10R \times 10R \times 10R$ with $R = 30$, which allow us to identify the influence of the wall confinement on the droplet migration. Periodic boundary conditions are applied on the upper and lower boundaries, while the half-way bounce-back boundary conditions [57] are applied in other boundaries. The fluid properties are set to $\rho^{R0} = \rho^{B0} = 1$, $\mu^{R0} = \mu^{B0} = 0.1$ and $\sigma = 0.001$. The droplet is covered with insoluble surfactants, with the concentration following the distribution given by Eq.(34) during the simulation. The surfactant related parameters are selected as $E_0 = 0.5$, $D_s = 0.1$ and $x_{in} = 0.5$.

Figure 4(a) plots the steady-state velocity field around the moving droplet for $L_x \times L_y \times L_z = 10R \times 10R \times 10R$. It is clear that under the action of Marangoni force caused by the surfactant gradient, the flow pattern within the droplet exhibits recirculation flows that make the droplet migrate along the direction of surfactant gradient. Figure 4(b) presents the evolution of the dimensionless velocity $U/U_{YGB}$ for two different domains, in which the droplet velocity $U$ is calculated by

$$U = \sum_{i,j} u_z(i,j) N(\rho^N(i,j)) / \sum_{i,j} N(\rho^N(i,j)) \tag{33}$$

with $N(\rho^N(i,j))$ defined by

$$N(\rho^N(i,j)) = \begin{cases} 0, & \rho^N(i,j) < 0 \\ 1, & \rho^N(i,j) \geq 0 \end{cases}, \tag{34}$$

and the dimensionless time is defined as $\tau = Ut/R$. It can be seen that $U$ first increases dramatically with time and then reaches a steady value in each domain. For the domain of $L_x \times L_y \times L_z = 5R \times 5R \times 10R$, $U/U_{YGB}$ finally converges to 0.958 only due to the wall confinement in both x and y directions; while as the domain size is increased to $10R \times 10R \times 10R$, $U/U_{YGB}$ becomes 0.993, showing good agreement with the theoretical prediction.

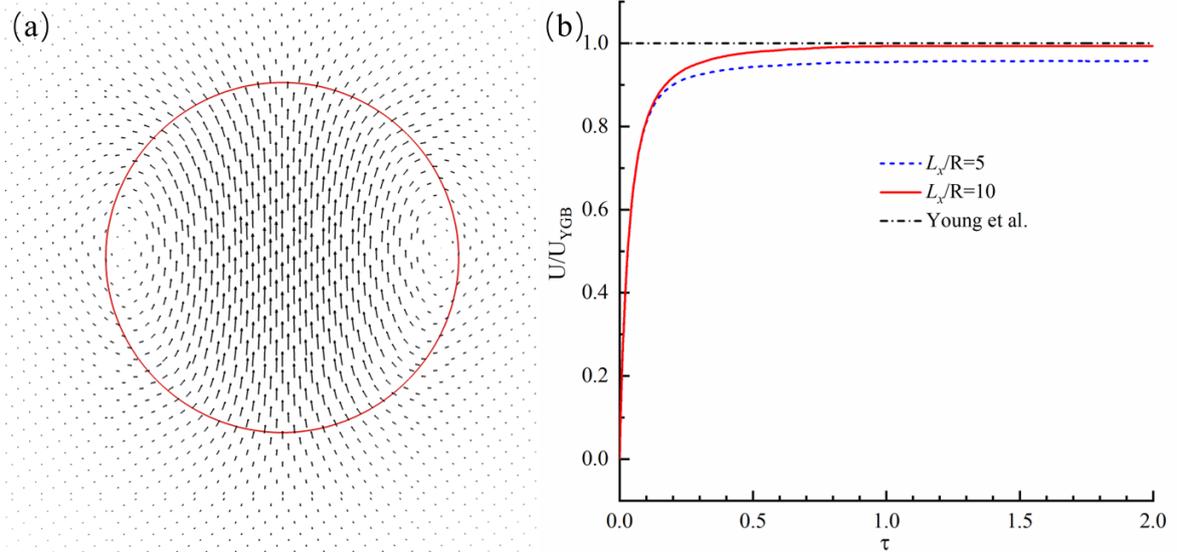

Figure 4 (a) Velocity field around the moving droplet in the *x-z* meridian plane for $L_x = L_y = 10R$, and the velocity vectors are plotted every 2 grid points for better visualization; (b) The evolution of the droplet velocity $U/U_{YGB}$ in two different domain sizes. In (b), the dash-dot lines represent the theoretical prediction from Eq.(35).

### *3.3 Deformation of a surfactant-covered droplet under simple shear*

Then, we perform the 2D simulations of a surfactant-covered droplet under simple shear flow, and compare the results of the droplet deformation with those obtained by [24]. As illustrated in Figure 5, a circular droplet (red fluid) with radius $R$ is initially placed halfway between two parallel walls that are separated by a distance $H$. The upper and lower walls move with opposite velocities of $u_w$ and $-u_w$ that produce a constant shear rate $\dot{\gamma} = 2u_w/H$. The length and velocity scales of this problem are taken as $L = R$ and $U = u_w/2$, and the droplet behavior can be characterized by three groups of dimensionless parameters: (1) the flow parameters including the capillary number $Ca = \mu^{B0}U/\sigma_0$, the Reynolds number $Re = \rho^{B0}UL/\mu^{B0}$, and the viscosity ratio $\lambda = \mu^{R0}/\mu^{B0}$; (2) the interface surfactant parameters including the surface Peclet number $Pe_s = UL/D_s$, the elasticity number $E_0 = RT\psi_\infty/\sigma_0$, and the surfactant coverage $x_{in} = \psi_0/\psi_\infty$; and (3) the bulk surfactant parameters including the Biot number $Bi = r_d/\dot{\gamma}$, the bulk Peclet number $Pe_b = UL/D_b$, the adsorption number $k = r_a\phi_0/r_d$, and the adsorption depth $h = \psi_0/(\phi_0 R)$.

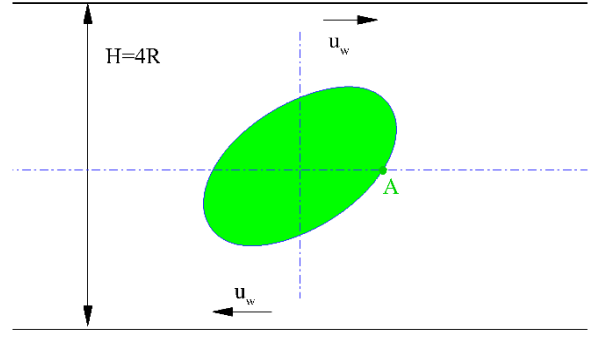

Figure 5 Illustration of a 2D droplet in a simple shear flow.

To facilitate the quantitative comparison, we conduct the simulations with the same parameters as those in Teigen et al. [24], i.e., $Pe_s = 10$, $\text{Re} = 1$, $Ca = 0.5$, $h = 0.5$ and $Bi = 1$ for the soluble surfactant cases with $Pe_b$ =0.1, 1 and 10, and the insoluble surfactant case. The Langmuir EOS is employed with $\sigma_0 = 5 \times 10^{-4}$, $x_{in} = 0.5$ and $E_0 = 0.3$. It is assumed that initially, the interface and bulk surfactants are distributed evenly, and there is no mass transfer across the interface, i.e, $j_0 = r_a \phi_0 (\psi_\infty - \psi_0) - r_d \psi_0 = 0$, leading to $k = x_{in}/(1-x_{in}) = 1$. The computational domain is set as $[-6R, 6R] \times [-2R, 2R]$ with $R = 50$ lattices. The half-way bounce-back boundary conditions [57] are applied on the upper and lower walls with opposite velocities, while periodic conditions are used in the horizontal direction. The dimensionless time is defined as $\tau = t\dot{\gamma}$, and the $x$ and $y$ coordinates are both normalized by the droplet radius $R$.

Figure 6 shows the droplet shapes at $\tau = 0$, 4 and 8 for the soluble surfactant cases with different values of $Pe_b$ and the insoluble surfactant case, where the solid lines represent the present results and the dashed lines represent the results of Teigen et al. [24]. It is observed that the droplet shapes obtained from the present method all agree well with those by Teigen et al. [24]. At $\tau = 4$ and 8, the droplet deforms into a more elongated shape in the soluble surfactant case than in the insoluble case, because when the droplet elongates, the bulk surfactants replenish the interface surfactants through adsorption, which reduces the interfacial tension. As the time passes by, the dilution of surfactants becomes more severe, promoting the adsorption onto the interface in the soluble surfactant cases, and thus the difference between the droplet shapes in the soluble and insoluble surfactant cases becomes more pronounced at $\tau = 8$. It is also observed that in the soluble cases, as $Pe_b$ increases, the droplet deformation gets smaller, which can be explained as follows. As $Pe_b$ increases, the bulk surfactants become increasingly non-uniform due to weaker bulk diffusion, as shown in Figure 7(e), which prevents further adsorption from bulk to interface, leading to a decrease of the interface surfactant concentration and thus to the decrease of droplet deformation.

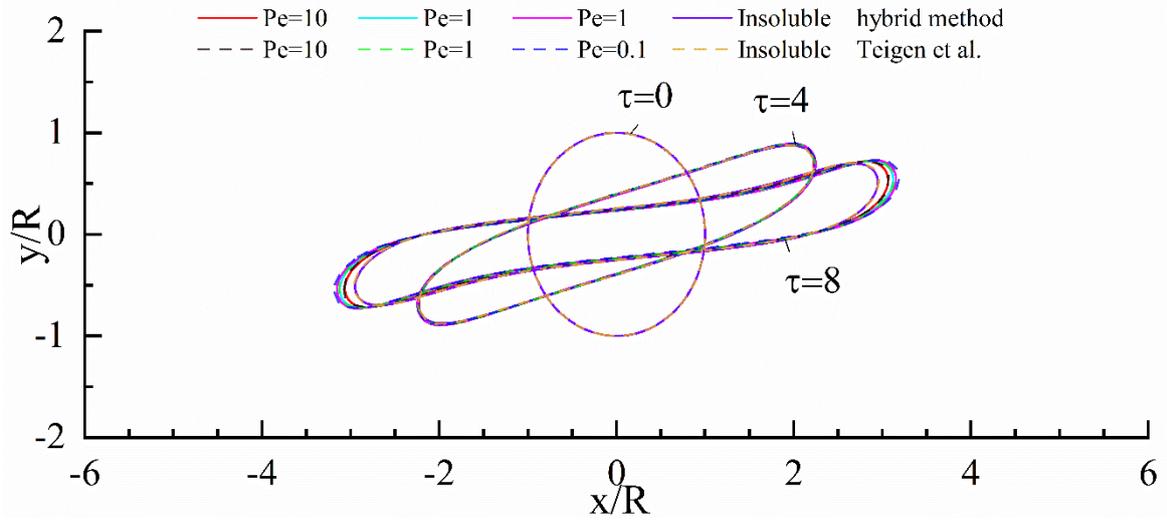

Figure 6 The droplet shapes at $\tau = 0$, 4 and 8 in the insoluble surfactant case and the soluble surfactant cases with different $Pe_b$. The solid lines represent the present results, while the dashed lines represent the results of Teigen et al. [24].

Figure 7(a) and (b) plot the distributions of the dimensionless interface surfactant concentration $\psi^* = \psi/\psi_0$ and interfacial tension $\sigma^* = \sigma/\sigma_0$ along the arc length $s$ at $\tau = 4$, together with the results obtained by Teigen et al. [24]. Herein, we measure the arc length $s$ counter-clockwise from the right intersection point between the interface (represented by $\rho^N = 0$) and the x-axis (point 'A' in Figure 5), and normalize it by $R$. It is seen that for all the cases considered, $\psi^*$ exhibits the same distributions in trend, with the highest $\psi^*$ value positioned at the droplet poles and the lowest value at the equator, while $\sigma^*$ changes in an opposite trend to $\psi^*$. In the insoluble surfactant case, $\psi^*$ is the most non-uniform among all the cases, and its value is lower than that in the soluble cases for most part of the interface, which leads to a lower droplet deformation. In the soluble surfactant cases, with the increase of $Pe_b$, $\psi^*$ decreases at most part of the droplet interface, and thus the average $\psi^*$ decreases, resulting in a decreased droplet deformation, as shown in Figure 5. In addition, we notice that regardless of the solubility and diffusivity of surfactants, all the obtained results match perfectly with the results of Teigen et al. [24].

The distributions of the dimensionless capillary force $F_{Ca}$ and Marangoni force $F_{Ma}$ along $s$ are plotted in Figure 7(c) and (d), where $F_{Ca}$ and $F_{Ma}$ are defined as $\kappa\sigma^*R$ and $\nabla_s\sigma^*\cdot\mathbf{\tau}R$ with $\mathbf{\tau}=(-n_y,n_x)$ being the unit vector tangential to the interface. Again, for all the cases considered, $F_{Ca}$ and $F_{Ma}$ agree well with those obtained by Teigen et al. [24]. It is also observed that $F_{Ca}$ achieves its maximum at the poles of the droplet, and the variation of $F_{Ca}$ for different $Pe_b$ is small, since $F_{Ca}$ is mainly determined by the interface curvature, which varies slightly except at the droplet poles. On the other hand, $Pe_b$ has a visible influence on the Marangoni force. $F_{Ma}$ is the most non-uniform in the insoluble surfactant case, and in the soluble cases, the $F_{Ma}$ distribution becomes more uniform as $Pe_b$ decreases.

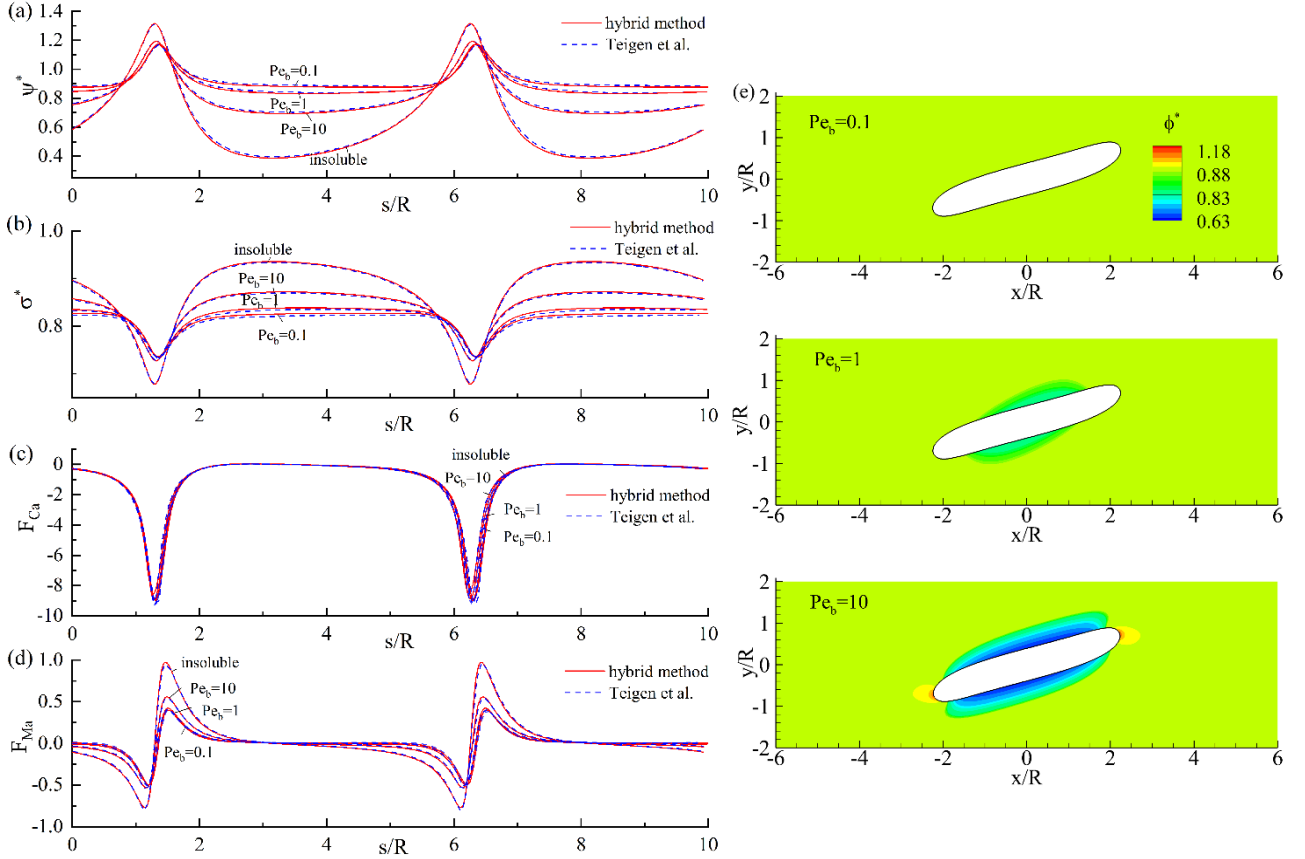

Figure 7 (a) The distributions of the dimensionless interface surfactant concentration $\psi^*$, (b) interfacial tension $\sigma^*$, (c) capillary force $F_{Ca}$ and (d) Marangoni force $F_{Ma}$ along the arc length $s$ at $\tau = 4$; (e) the distributions of the bulk surfactant concentration $\phi^*$ for $Pe_b$=0.1, 1 and 10 at $\tau = 4$. In (a)-(d), the solid lines represent the present results, while the dashed lines represent the results of Teigen et al. [24].

### *3.4 Buoyancy-driven rising bubbles*

Finally, we study the bubble rise driven by the buoyancy force in a surfactant solution, in which the effect of surfactants is identified for varying geometrical confinement, Eotvos number, and Biot number. A spherical bubble (red fluid) with radius $R$, immersed in a liquid (blue fluid) containing soluble surfactants, is initially located at the centerline of the channel near the bottom boundary, and the channel has a size of $L_x \times L_y \times L_z$. Periodic boundary conditions are applied on the top and bottom boundaries, while no-slip boundary conditions are applied on other boundaries, with the implementation using the half-way bounce-back scheme. An additional buoyancy force $\mathbf{F}_b = (\rho - \rho^{B0})g\mathbf{k}$ is introduced to the LBM in the same way as the interfacial force $\mathbf{F}_s$, where $g$ is the gravitational acceleration, and $\mathbf{k}$ is the unit vector in the $z$ direction. Unlike the droplet deformation under shear flow, three important dimensionless parameters, namely the Eotvos number $Eo = \Delta\rho g (2R)^2 / \sigma$, the Morton number $Mo = \Delta\rho g (\mu^{B0})^4 / (\rho^{B0})^2 \sigma^3$ and the confinement ratio $2R/L_x$, are often used to characterize the bubble rise in a confined channel, in which $\Delta\rho = \rho^{B0} - \rho^{R0}$ is the density difference. For the sake of simplicity, the fluid properties are set to $\rho^{R0} = 0.1$, $\rho^{B0} = 1$, $\mu^{R0}/\mu^{B0} = 0.1$ and $\sigma = 0.004$. Unless otherwise stated, the surfactant parameters are selected as $E_0 = 0.5$, $Pe_b = Pe_s = 10$, $k=1$, $h=0.5$ and $Bi=10$, and the bubble interface is initially clean with $x_{in} = 0$.

The simulations are first conducted with $Eo=1$ and $Mo=0.1$, under which the bubble eventually attains a nearly spherical shape. For a clean spherical bubble in an infinite domain, its terminal velocity $U_{th}$ can be given by Hadamard-Rybczynski solution [58] as

$$U_{th} = \frac{2}{3}\frac{g\Delta\rho R^2}{\mu^{B0}}\frac{\mu^{B0}+\mu^{R0}}{2\mu^{B0}+3\mu^{R0}}. \tag{35}$$

Three different domain sizes, i.e. $L_x \times L_y \times L_z = 5R \times 5R \times 10R$, $10R \times 10R \times 10R$ and $20R \times 20R \times 10R$, are considered, corresponding to $2R/L_x = 0.4$, 0.2 and 0.1, respectively. Figure 8 plots the evolution of the dimensionless bubble velocity, defined as the bubble velocity $U$ divided by $U_{th}$, in the clean and surfactant cases. We can see that in the clean case, $U/U_{th}$ increases continuously until reaching a steady-state value for each confinement ratio, and the steady-state $U/U_{th}$ values are 0.552, 0.811 and 0.937 for $2R/L_x = 0.4$, 0.2 and 0.1, respectively. Evidently, the terminal velocity of bubble becomes closer to the Hadamard-Rybczynski solution $U_{th}$ as the confinement ratio decreases, but it is always lower than $U_{th}$, which suggests that the effect of wall confinement cannot be totally neglected even for the confinement ratio as low as 0.1, consistent with the previous results [16, 58]. Quantitatively, the terminal velocity ($U/U_{th} = 0.937$) obtained at $2R/L_x = 0.1$ also shows excellent agreement with the one ($U/U_{th} = 0.9303$) obtained by Metin et al. [16]. Unlike a bubble in the clean case, $U/U_{th}$ first increases rapidly, then decreases, and finally reaches a steady-state value in the surfactant case. In addition, the steady-state value is lower than that in the corresponding clean case, indicating the retardation effect of surfactants, which was also found by Clift et al. [58] and Metin et al. [16]. To explain such retardation effect, Figure 9 shows the typical distribution of interface and bulk surfactant concentrations in the surfactant case where $2R/L_x = 0.4$. It is seen that surfactants are swept to the rear of the bubble, where they accumulate, giving rise to a local decrease in interfacial tension. The resulting interfacial tension gradient directed from rear to front results in a Marangoni stress that resists interface motion, thus rigidifying the bubble interface. This can be further identified in Figure 10, where we show the steady-state velocity fields in a frame moving with the bubble for the clean and surfactant cases with $2R/L_x = 0.4$. A pair of vortices is generated inside the bubble for the clean case, while the vortices inside the bubble nearly vanish for the surfactant case, which demonstrates the inhibition effect of surfactants in the interface motion caused by the Marangoni stress.

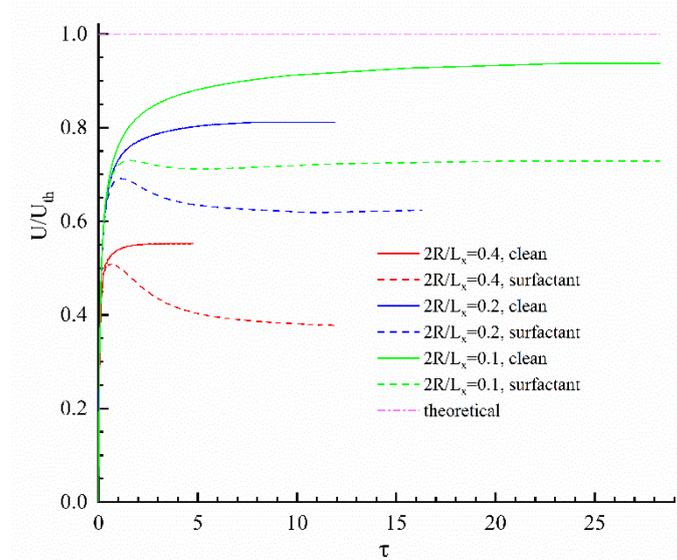

Figure 8 Time evolution of the dimensionless bubble rising velocity $U/U_{th}$ for different confinement ratios. The dimensionless time is defined as $\tau = t/\sqrt{2R/g}$.

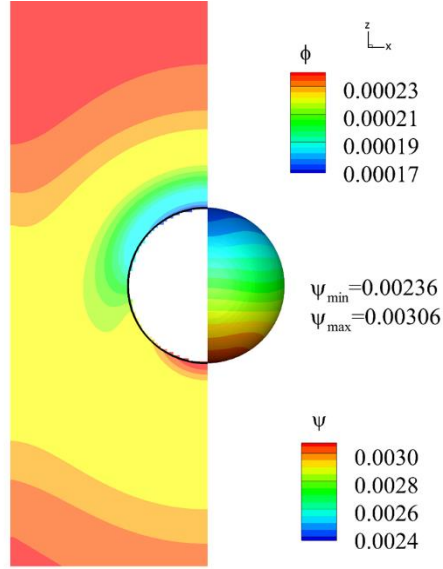

Figure 9 The distributions of the bulk surfactant concentration $\phi$ and the interface surfactant concentration $\psi$ at the interface in the x-z mid-plane for the confinement ratio of 0.4.

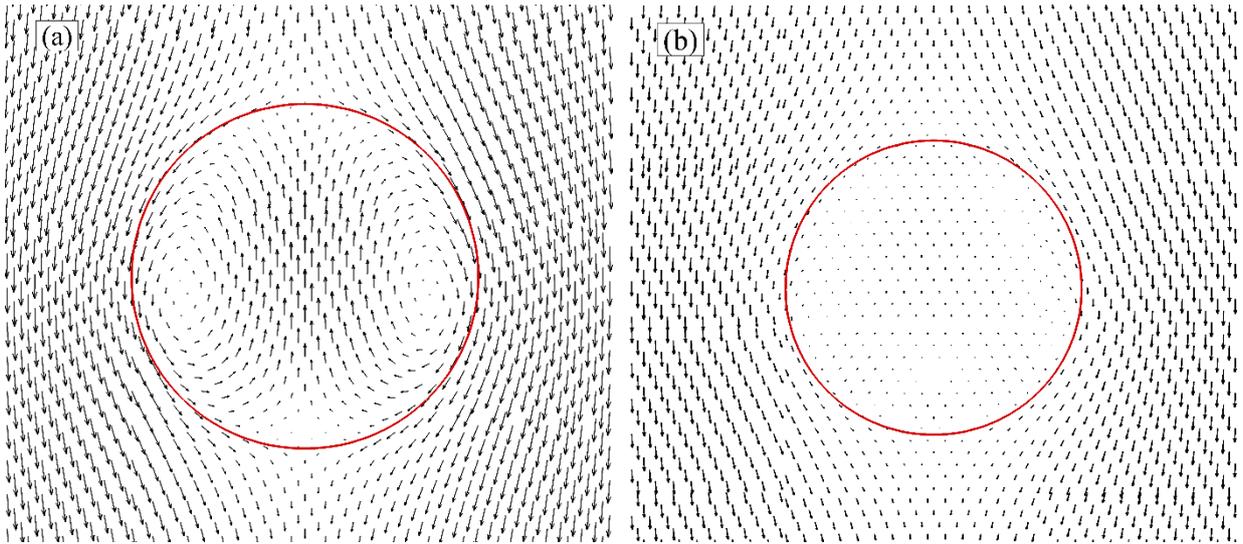

Figure 10 The steady-state velocity fields in a coordinate system moving with the bubble for (a) clean and (b) surfactant cases with $2R/L_x = 0.4$. The velocity vectors are plotted every 3 grid points for better visualization.

Then, we choose the confinement ratio $2R/L_x = 0.4$ and increase the Eotvos number to $Eo = 10$, at which the bubble no longer exhibits a spherical shape. All the parameters are kept the same as those in $Eo = 1$ simulations except $Bi$, which is varies from 0.5 to 10 to explore the effect of the mass transfer rate of surfactants. Since Eq.(38) cannot be used to predict the terminal velocity of a deformable bubble, it would not be a good option to use $U/U_{th}$ to describe the bubble rise. Following the previous work [15, 40], we track the evolution of the Reynolds number $Re = 2\rho^{B0}UR/\mu^{B0}$ instead of $U/U_{th}$. Figure 11 presents the evolution of $Re$ in the clean case and in the surfactant cases with $Bi=0.5$, 1 and 10. It is seen that in the clean case, $Re$ increases until reaching a steady-state value. Whereas in the surfactant cases, the final Re is reduced due to the retardation effect of surfactants, but the influence of surfactants is not as significant as when $Eo=1$. This is because surfactants take effect by modifying the interfacial tension, and thus the influence of surfactants is more pronounced at a higher interfacial tension, which corresponds to a lower $Eo$. In Figure 12, we find that in the clean and surfactant cases, a pair of vortices always appears inside the bubble, but the velocities are smaller in the surfactant cases. In addition, as $Bi$ increases, the final $Re$ increases and the retardation effect of surfactants weakens. To clarify its cause, we plot the evolution of the average, minimum and maximum interface surfactant concentrations ($\psi_{av}$, $\psi_{min}$ and $\psi_{max}$) for different $Bi$ in Figure 13. It is found that the final $\psi_{av}$ values are almost the same for various $Bi$, indicating that the average interfacial tension does not contribute to the increase of $Re$ with $Bi$, while the increasing rate of $\psi_{av}$ rises with $Bi$ due to the faster surfactant adsorption and desorption. On the other hand, as $Bi$ increases, the steady-state $\psi_{min}$ increases but $\psi_{max}$ decreases, leading to the decrease of non-uniform effects and thus the Marangoni stresses, which weaken the retardation effect of surfactants.

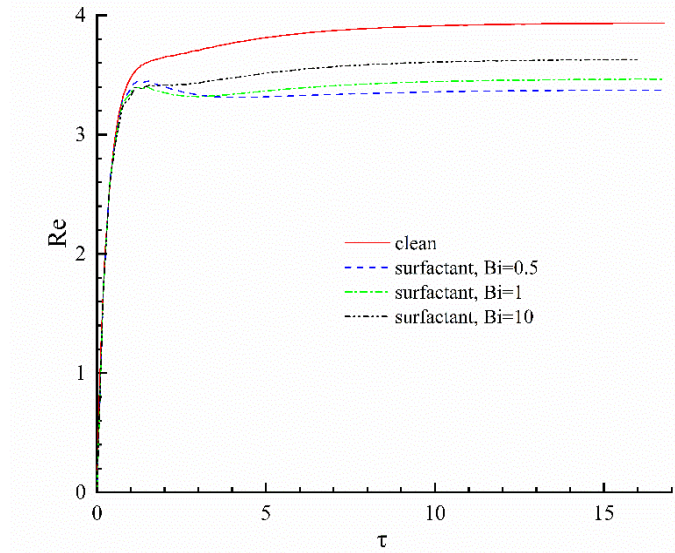

Figure 11 Time evolution of $Re$ in the clean case and the surfactant cases with different Biot numbers for $2R/L_x=0.4$. The dimensionless time is defined as $\tau=t/\sqrt{2R/g}$.

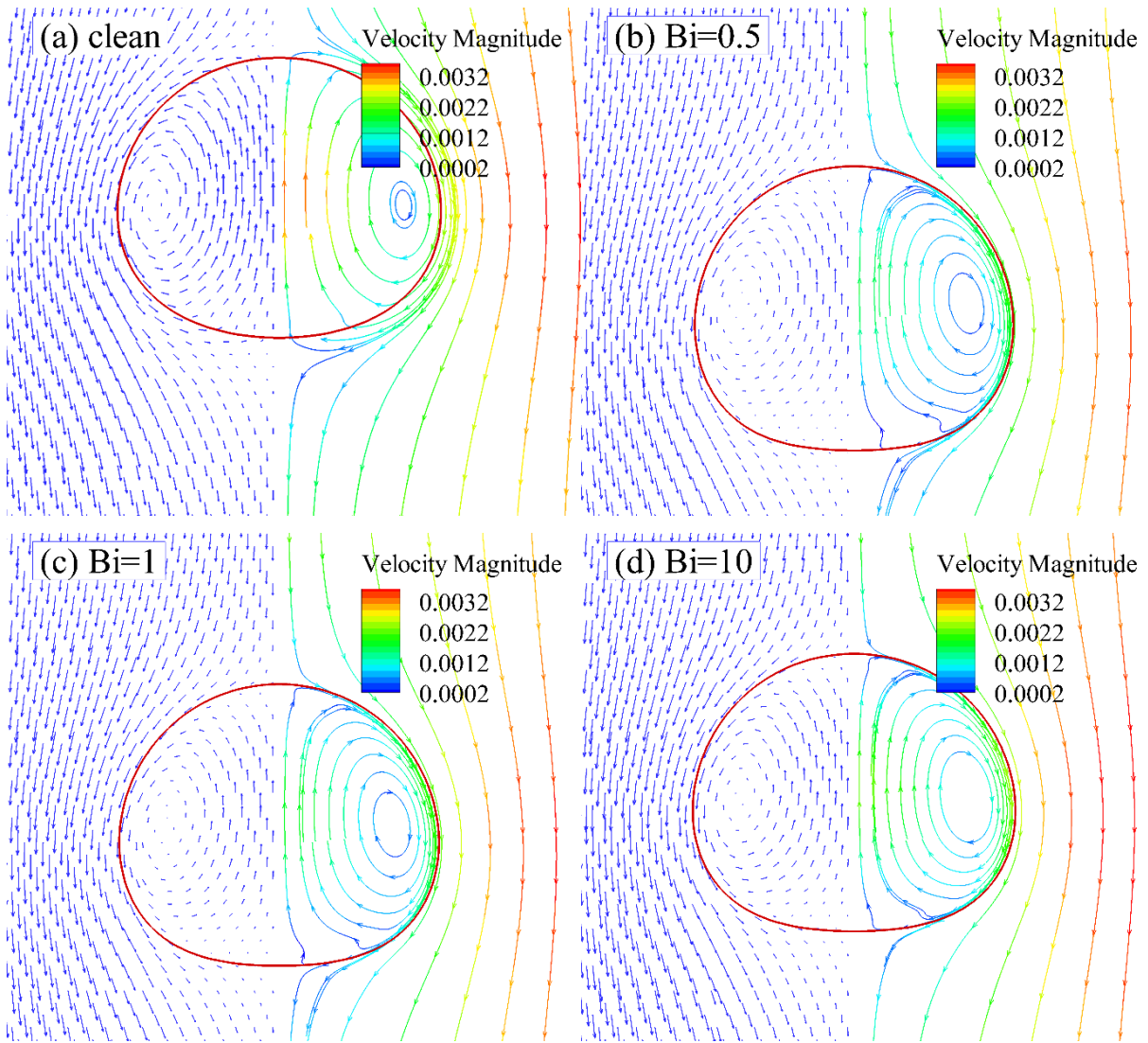

Figure 12 The steady-state velocity vectors (left) and streamlines (right) in a coordinate system moving with the bubble for the (a) clean case and surfactant cases with (b) $Bi=0.5$, (c) $Bi=1$ and (d) $Bi=10$.

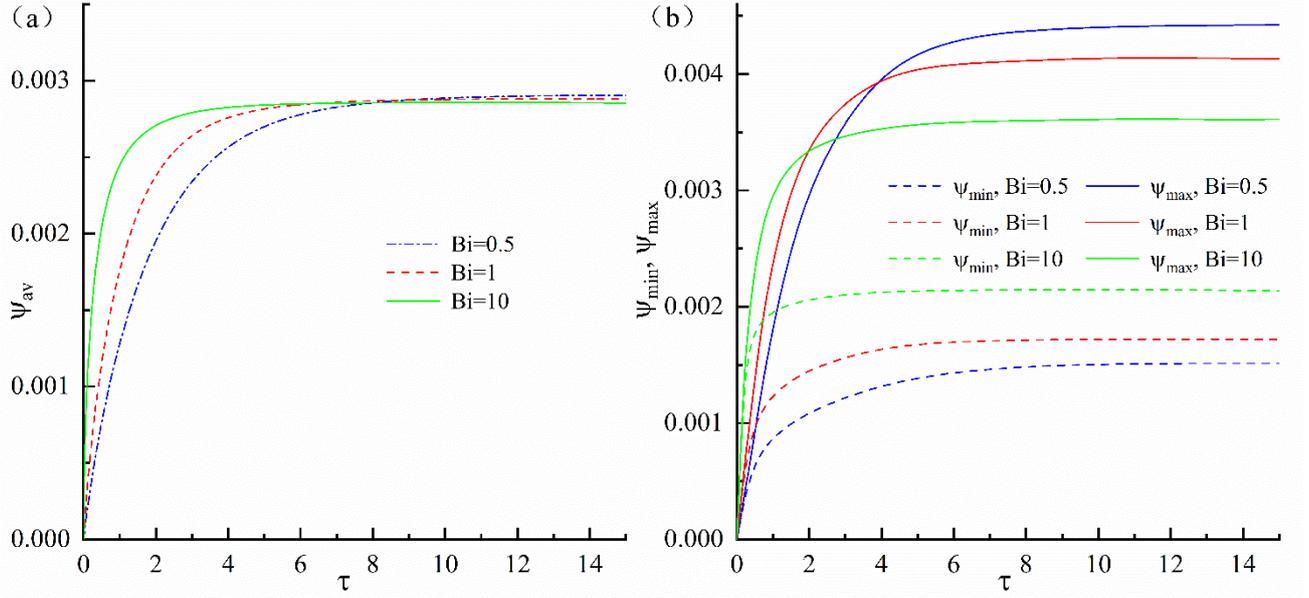

Figure 13 Time evolution of (a) the average interface surfactant concentration $\psi_{av}$, and (b) the minimum and maximum interface surfactant concentrations for different values of $Bi$. The dimensionless time is defined as $\tau = t/\sqrt{2R/g}$.

## 4 Conclusions

In this work, we develop a hybrid LB-FD method to simulate interfacial flows with soluble surfactants. The convection-diffusion equations governing the interface and bulk surfactant concentrations, expressed in a diffuse-interface form, are solved by a FD method, in which the source terms are added to account for the adsorption and desorption dynamics between the interface and bulk surfactants, while the two-phase flow is solved by an improved LB color-gradient model, which enables the hybrid method to simulate binary fluids with unequal densities. The flow field and the interface surfactant concentration field are coupled through an equation of state, which however is not limited to the Langmuir one. The capability and accuracy of the hybrid method are first validated by simulating three typical cases, namely the adsorption of bulk surfactants onto the interface of a stationary droplet, the droplet migration in a constant surfactant gradient, and the deformation of a surfactant-covered droplet in a simple shear flow, in which the simulated results are quantitatively compared with the theoretical solutions and/or available literature data. Then, the hybrid method is used to study the bubble rise in a liquid containing soluble surfactants. It is found that at $Eo=1$, the terminal velocity of the bubble decreases with confinement ratio in either clean or surfactant case, and the presence of surfactants has an obvious retardation effect on the bubble rise due to the Marangoni stress that resists interface motion. Increasing $Eo$ weakens the influence of surfactants, and by varying $Bi$ at $Eo=10$, we show that as $Bi$ increases, the non-uniform effects of surfactants and thus the Marangoni stresses decrease, leading to a reduced retardation effect on the bubble rise.

**Acknowledgements**

This work is supported by the National Natural Science Foundation of China (Grant Nos. 51906206, 12072257, 51976174), the Natural Science Foundation of Shaanxi Province of China (Grant No. 2020JQ-191), and the Major Special Science and Technology Project of the Inner Mongolia Autonomous Region (Grant No. 2020ZD0022).

**Data availability**

Data will be made available on request.

**Appendix A: Transformation matrices, equilibrium distributions and correction terms in moment space and diagonal relaxation matrices in 2D and 3D**

The transformation matrix $\mathbf{M}$, the equilibrium distributions $\mathbf{m}^{k,eq}$, the correction term $\mathbf{C}^k$ and the diagonal relaxation matrix are given by [42, 45]

$$\mathbf{M}=\begin{bmatrix} 1 & 1 & 1 & 1 & 1 & 1 & 1 & 1 & 1 \\ -4 & -1 & -1 & -1 & -1 & 2 & 2 & 2 & 2 \\ 4 & -2 & -2 & -2 & -2 & 1 & 1 & 1 & 1 \\ 0 & 1 & 0 & -1 & 0 & 1 & -1 & -1 & 1 \\ 0 & -2 & 0 & 2 & 0 & 1 & -1 & -1 & 1 \\ 0 & 0 & 1 & 0 & -1 & 1 & 1 & -1 & -1 \\ 0 & 0 & -2 & 0 & 2 & 1 & 1 & -1 & -1 \\ 0 & 1 & -1 & 1 & -1 & 0 & 0 & 0 & 0 \\ 0 & 0 & 0 & 0 & 0 & 1 & -1 & 1 & -1 \end{bmatrix}, \qquad (36)$$

$$\mathbf{m}^{k,eq} = \begin{bmatrix} \rho^k \\ \rho^k\left(-3.6\alpha^k - 0.4 + 3\mathbf{u}^2\right) \\ \rho^k\left(5.4\alpha^k - 1.4 - 3\mathbf{u}^2\right) \\ \rho^k u_x \\ \rho^k\left(-1.8\alpha^k - 0.2\right)u_x \\ \rho^k u_y \\ \rho^k\left(-1.8\alpha^k - 0.2\right)u_y \\ \rho^k\left(u_x^2 - u_y^2\right) \\ \rho^k u_x u_y \end{bmatrix}, \quad \mathbf{C}^k = \begin{bmatrix} 0 \\ 3Q_x + 3Q_y \\ 0 \\ 0 \\ 0 \\ 0 \\ 0 \\ Q_x - Q_y \\ 0 \end{bmatrix}, \tag{37}$$

$$\mathbf{S} = diag\left(0, 1.25, 1.14, 0, 1.6, 0, 1.6, s_v, s_v\right) \tag{38}$$

in 2D, and

$$\mathbf{M} = \begin{bmatrix} 1 & 1 & 1 & 1 & 1 & 1 & 1 & 1 & 1 & 1 & 1 & 1 & 1 & 1 & 1 & 1 & 1 & 1 & 1 \\ 0 & 1 & -1 & 0 & 0 & 0 & 0 & 1 & -1 & 1 & -1 & 1 & -1 & 1 & -1 & 0 & 0 & 0 & 0 \\ 0 & 0 & 0 & 1 & -1 & 0 & 0 & 1 & -1 & -1 & 1 & 0 & 0 & 0 & 0 & 1 & -1 & 1 & -1 \\ 0 & 0 & 0 & 0 & 0 & 1 & -1 & 0 & 0 & 0 & 0 & 1 & -1 & -1 & 1 & 1 & -1 & -1 & 1 \\ 0 & 1 & 1 & 1 & 1 & 1 & 1 & 2 & 2 & 2 & 2 & 2 & 2 & 2 & 2 & 2 & 2 & 2 & 2 \\ 0 & 2 & 2 & -1 & -1 & -1 & -1 & 1 & 1 & 1 & 1 & 1 & 1 & 1 & 1 & -2 & -2 & -2 & -2 \\ 0 & 0 & 0 & 1 & 1 & -1 & -1 & 1 & 1 & 1 & 1 & -1 & -1 & -1 & -1 & 0 & 0 & 0 & 0 \\ 0 & 0 & 0 & 0 & 0 & 0 & 0 & 1 & 1 & -1 & -1 & 0 & 0 & 0 & 0 & 0 & 0 & 0 & 0 \\ 0 & 0 & 0 & 0 & 0 & 0 & 0 & 0 & 0 & 0 & 0 & 1 & 1 & -1 & -1 & 0 & 0 & 0 & 0 \\ 0 & 0 & 0 & 0 & 0 & 0 & 0 & 0 & 0 & 0 & 0 & 0 & 0 & 0 & 0 & 1 & 1 & -1 & -1 \\ 0 & 0 & 0 & 0 & 0 & 0 & 0 & 1 & -1 & -1 & 1 & 0 & 0 & 0 & 0 & 0 & 0 & 0 & 0 \\ 0 & 0 & 0 & 0 & 0 & 0 & 0 & 1 & -1 & 1 & -1 & 0 & 0 & 0 & 0 & 0 & 0 & 0 & 0 \\ 0 & 0 & 0 & 0 & 0 & 0 & 0 & 0 & 0 & 0 & 0 & 1 & -1 & -1 & 1 & 0 & 0 & 0 & 0 \\ 0 & 0 & 0 & 0 & 0 & 0 & 0 & 0 & 0 & 0 & 0 & 1 & -1 & 1 & -1 & 0 & 0 & 0 & 0 \\ 0 & 0 & 0 & 0 & 0 & 0 & 0 & 0 & 0 & 0 & 0 & 0 & 0 & 0 & 0 & 1 & -1 & -1 & 1 \\ 0 & 0 & 0 & 0 & 0 & 0 & 0 & 0 & 0 & 0 & 0 & 0 & 0 & 0 & 0 & 1 & -1 & 1 & -1 \\ 0 & 0 & 0 & 0 & 0 & 0 & 0 & 1 & 1 & 1 & 1 & 0 & 0 & 0 & 0 & 0 & 0 & 0 & 0 \\ 0 & 0 & 0 & 0 & 0 & 0 & 0 & 0 & 0 & 0 & 0 & 1 & 1 & 1 & 1 & 0 & 0 & 0 & 0 \\ 0 & 0 & 0 & 0 & 0 & 0 & 0 & 0 & 0 & 0 & 0 & 0 & 0 & 0 & 0 & 1 & 1 & 1 & 1 \end{bmatrix}, \tag{39}$$

$$\mathbf{m}^{k,eq} = \begin{bmatrix} \rho^k \\ \rho^k u_x \\ \rho^k u_y \\ \rho^k u_z \\ 3p^k + \rho^k |\mathbf{u}|^2 \\ \rho^k (2u_x^2 - u_y^2 - u_z^2) \\ \rho^k (u_y^2 - u_z^2) \\ \rho^k u_x u_y \\ \rho^k u_x u_z \\ \rho^k u_y u_z \\ p^k u_y \\ p^k u_x \\ p^k u_z \\ p^k u_x \\ p^k u_z \\ p^k u_y \\ \rho^k \left[ \left(1 - \alpha^k - |\mathbf{u}|^2\right) + 2\left(u_x^2 + u_y^2\right) \right]/6 \\ \rho^k \left[ \left(1 - \alpha^k - |\mathbf{u}|^2\right) + 2\left(u_x^2 + u_z^2\right) \right]/6 \\ \rho^k \left[ \left(1 - \alpha^k - |\mathbf{u}|^2\right) + 2\left(u_y^2 + u_z^2\right) \right]/6 \end{bmatrix}, \quad \mathbf{C}^k = \begin{bmatrix} 0 \\ 0 \\ 0 \\ 0 \\ Q_x + Q_y + Q_z \\ 2Q_x - Q_y - Q_z \\ Q_y - Q_z \\ 0 \\ 0 \\ 0 \\ 0 \\ 0 \\ 0 \\ 0 \\ 0 \\ 0 \\ 0 \\ 0 \\ 0 \end{bmatrix}, \quad (40)$$

$$\mathbf{S} = diag\left(1,1,1,1,1,s_v,s_v,s_v,s_v,s_v,1,1,1,1,1,1,1,1,1\right) \quad (41)$$

in 3D.

In the above equations, $\alpha^k$ is a free parameter related to the speed of sound of the fluid $k$, $c_s^k$, by $\left(c_s^k\right)^2 = \frac{3}{5}\left(1-\alpha^k\right)$ in 2D and $\left(c_s^k\right)^2 = 0.5\left(1-\alpha^k\right)$ in 3D; $Q_x$, $Q_y$, and $Q_z$ if present, are taken as $Q_x = \left(1.8\alpha^k - 0.8\right)\partial_x\left(\rho^k u_x\right)$ and $Q_y = \left(1.8\alpha^k - 0.8\right)\partial_y\left(\rho^k u_y\right)$ in 2D, and $Q_x = \partial_x\left\{\rho^k u_x\left[c^2 - \left(c_s^k\right)^2\right]\right\}$, $Q_y = \partial_y\left\{\rho^k u_y\left[c^2 - \left(c_s^k\right)^2\right]\right\}$ and $Q_z = \partial_z\left\{\rho^k u_z\left[c^2 - \left(c_s^k\right)^2\right]\right\}$ in 3D. To obtain a stable interface, the free parameters $\alpha^R$ and $\alpha^B$ should satisfy [59]

$$\gamma = \frac{\rho^{R0}}{\rho^{B0}} = \frac{1-\alpha^B}{1-\alpha^R}, \quad (42)$$

where $\gamma$ represents the density ratio of red fluid to blue fluid. In the diagonal relaxation matrix $\mathbf{S}$, $s_v$ is related to the dynamic viscosity of the mixture fluid, $\mu$, by $\mu = \delta_t \left( \dfrac{1}{s_v} - \dfrac{1}{2} \right) \sum_k \rho^k \left( c_s^k \right)^2$, where $\mu$ can be computed using the viscosity of the pure red fluid ($\mu^{R0}$) and the viscosity of the pure blue fluid ($\mu^{B0}$) as

$$\frac{1}{\mu} = \frac{1+\rho^N}{2\mu^{R0}} + \frac{1-\rho^N}{2\mu^{B0}}. \tag{43}$$

**References**

[1] B. Riechers, F. Maes, E. Akoury, B. Semin, P. Gruner, J.C. Baret, Surfactant adsorption kinetics in microfluidics, Proceedings of the National Academy of Sciences of the United States of America, 113 (2016) 11465-11470.
[2] J.-C. Baret, Surfactants in droplet-based microfluidics, Lab on a Chip, 12 (2012) 422-433.
[3] M.P.N. Bui, C.A. Li, K.N. Han, J. Choo, E.K. Lee, G.H. Seong, Enzyme Kinetic Measurements Using a Droplet-Based Microfluidic System with a Concentration Gradient, Anal. Chem., 83 (2011) 1603-1608.
[4] S. De, S. Malik, A. Ghosh, R. Saha, B. Saha, A review on natural surfactants, Rsc Advances, 5 (2015) 65757-65767.
[5] S.L. Anna, N. Bontoux, H.A. Stone, Formation of dispersions using "flow focusing" in microchannels, Applied Physics Letters, 82 (2003) 364-366.
[6] G.T. Vladisavljevic, R. Al Nuumani, S.A. Nabavi, Microfluidic Production of Multiple Emulsions, Micromachines, 8 (2017) 75.
[7] H. Liu, Y. Ba, L. Wu, Z. Li, G. Xi, Y. Zhang, A hybrid lattice Boltzmann and finite difference method for droplet dynamics with insoluble surfactants, Journal of Fluid Mechanics, 837 (2018) 381-412.
[8] H.A. Stone, A simple derivation of the time-dependent convective-diffusion equation for surfactant transport along a deforming interface, Physics of Fluids A: Fluid Dynamics, 2 (1990) 111-112.
[9] C.D. Eggleton, Y.P. Pawar, K.J. Stebe, Insoluble surfactants on a drop in an extensional flow: a generalization of the stagnated surface limit to deforming interfaces, Journal of Fluid Mechanics, 385 (1999) 79-99.
[10] K. Feigl, D. Megias-Alguacil, P. Fischer, E.J. Windhab, Simulation and experiments of droplet deformation and orientation in simple shear flow with surfactants, Chemical engineering science, 62 (2007) 3242-3258.
[11] G. Tryggvason, B. Bunner, A. Esmaeeli, D. Juric, N. Al-Rawahi, W. Tauber, J. Han, S. Nas, Y.J. Jan, A Front-Tracking Method for the Computations of Multiphase Flow, Journal of Computational Physics, 169 (2001) 708-759.

[12] J.-J. Xu, Z. Li, J. Lowengrub, H. Zhao, A level-set method for interfacial flows with surfactant, Journal of Computational Physics, 212 (2006) 590-616.
[13] A.J. James, J. Lowengrub, A surfactant-conserving volume-of-fluid method for interfacial flows with insoluble surfactant, Journal of Computational Physics, 201 (2004) 685-722.
[14] W.J. Milliken, L.G. Leal, The Influence of Surfactant on the Deformation and Breakup of a Viscous Drop: The Effect of Surfactant Solubility, Journal of Colloid and Interface Science, 166 (1994) 275-285.
[15] M. Muradoglu, G. Tryggvason, A front-tracking method for computation of interfacial flows with soluble surfactants, Journal of computational physics, 227 (2008) 2238-2262.
[16] M. Muradoglu, G. Tryggvason, Simulations of soluble surfactants in 3D multiphase flow, Journal of Computational Physics, 274 (2014) 737-757.
[17] S. Tasoglu, U. Demirci, M. Muradoglu, The effect of soluble surfactant on the transient motion of a buoyancy-driven bubble, Phys. Fluids, 20 (2008) 040805.
[18] S. Ganesan, L. Tobiska, Arbitrary Lagrangian-Eulerian finite-element method for computation of two-phase flows with soluble surfactants, Journal of Computational Physics, 231 (2012) 3685-3702.
[19] S. Shin, J. Chergui, D. Juric, L. Kahouadji, O.K. Matar, R.V. Craster, A hybrid interface tracking - level set technique for multiphase flow with soluble surfactant, Journal of Computational Physics, 359 (2018) 409-435.
[20] K.-Y. Chen, M.-C. Lai, A conservative scheme for solving coupled surface-bulk convection-diffusion equations with an application to interfacial flows with soluble surfactant, Journal of Computational Physics, 257 (2014) 1-18.
[21] S. Khatri, A.-K. Tornberg, An embedded boundary method for soluble surfactants with interface tracking for two-phase flows, Journal of Computational Physics, 256 (2014) 768-790.
[22] W.F. Hu, M.C. Lai, C. Misbah, A coupled immersed boundary and immersed interface method for interfacial flows with soluble surfactant, Comput. Fluids, 168 (2018) 201-215.
[23] H. Liu, Y. Zhang, Phase-field modeling droplet dynamics with soluble surfactants, Journal of Computational Physics, 229 (2010) 9166-9187.
[24] K.E. Teigen, P. Song, J. Lowengrub, A. Voigt, A diffuse-interface method for two-phase flows with soluble surfactants, Journal of Computational Physics, 230 (2011) 375-393.
[25] R.G.M. van der Sman, M.B.J. Meinders, Analysis of improved Lattice Boltzmann phase field method for soluble surfactants, Computer Physics Communications, 199 (2016) 12-21.
[26] G. Soligo, A. Roccon, A. Soldati, Breakage, coalescence and size distribution of surfactant-laden droplets in turbulent flow, Journal of Fluid Mechanics, 881 (2019) 244-282.
[27] G. Soligo, A. Roccon, A. Soldati, Coalescence of surfactant-laden drops by Phase Field Method, Journal of Computational Physics, 376 (2019) 1292-1311.
[28] G.P. Zhu, J.S. Kou, S.Y. Sun, J. Yao, A.F. Li, Numerical Approximation of a Phase-Field Surfactant Model with Fluid Flow, Journal of Scientific Computing, 80 (2019) 223-247.
[29] G.P. Zhu, J.S. Kou, J. Yao, A.F. Li, S.Y. Sun, A phase-field moving contact line model with soluble surfactants, Journal of Computational Physics, 405 (2020) 29109170.
[30] E.K. Far, M. Gorakifard, E. Fattahi, Multiphase Phase-Field Lattice Boltzmann Method for Simulation of Soluble Surfactants, Symmetry-Basel, 13 (2021) 1019.
[31] Y.J. Zong, C.H. Zhang, H. Liang, L. Wang, J.R. Xu, Modeling surfactant-laden droplet dynamics

by lattice Boltzmann method, Phys. Fluids, 32 (2020) 122105.
[32] J. Zhang, H. Liu, B. Wei, J. Hou, F. Jiang, Pore-Scale Modeling of Two-Phase Flows with Soluble Surfactants in Porous Media, Energy & Fuels, 35 (2021) 19374-19388.
[33] J. Zhang, X. Zhang, W. Zhao, H. Liu, Y. Jiang, Effect of surfactants on droplet generation in a microfluidic T-junction: A lattice Boltzmann study, Phys. Fluids, 34 (2022) 042121.
[34] Z.J. Tan, Y. Tian, J.X. Yang, Y.Y. Wu, J. Kim, An energy-stable method for a phase-field surfactant model, International Journal of Mechanical Sciences, 233 (2022) 107648.
[35] R.G.M. van der Sman, S. van der Graaf, Emulsion droplet deformation and breakup with Lattice Boltzmann model, Computer Physics Communications, 178 (2008) 492-504.
[36] T.Y. Zhang, Q. Wang, Cahn-Hilliard vs Singular Cahn-Hilliard Equations in Phase Field Modeling, Commun. Comput. Phys., 7 (2010) 362-382.
[37] J.-J. Xu, W. Shi, M.-C. Lai, A level-set method for two-phase flows with soluble surfactant, Journal of Computational Physics, 353 (2018) 336-355.
[38] A.K. Gunstensen, D.H. Rothman, S. Zaleski, G. Zanetti, Lattice Boltzmann model of immiscible fluids, Physical Review A, 43 (1991) 4320-4327.
[39] T. Reis, T.N. Phillips, Lattice Boltzmann model for simulating immiscible two-phase flows, Journal of Physics a: Mathematical and Theoretical, 40 (2007) 4033-4053.
[40] H. Liu, A.J. Valocchi, Q. Kang, Three-dimensional lattice Boltzmann model for immiscible two-phase flow simulations, Physical Review E, 85 (2012) 046309046309.
[41] S. Leclaire, M. Reggio, J.Y. Trepanier, Isotropic color gradient for simulating very high-density ratios with a two-phase flow lattice Boltzmann model, Comput. Fluids, 48 (2011) 98-112.
[42] Y. Ba, H. Liu, Q. Li, Q. Kang, J. Sun, Multiple-relaxation-time color-gradient lattice Boltzmann model for simulating two-phase flows with high density ratio, Physical Review E, 94 (2016) 023310.
[43] J. Zhang, H. Liu, Y. Ba, Numerical Study of Droplet Dynamics on a Solid Surface with Insoluble Surfactants, Langmuir, 35 (2019) 7858-7870.
[44] H. Liu, J. Zhang, Y. Ba, N. Wang, L. Wu, Modelling a surfactant-covered droplet on a solid surface in three-dimensional shear flow, Journal of Fluid Mechanics, 897 (2020) A33.
[45] Z.X. Wen, Q. Li, Y. Yu, K.H. Luo, Improved three-dimensional color-gradient lattice Boltzmann model for immiscible two-phase flows, Physical Review E, 100 (2019) 13023301.
[46] H. Nganguia, Y.N. Young, P.M. Vlahovska, J. Blawzdziewicz, J. Zhang, H. Lin, Equilibrium electro-deformation of a surfactant-laden viscous drop, Phys. Fluids, 25 (2013) 092106.
[47] Q. Li, K. Luo, Y. He, Y. Gao, W. Tao, Coupling lattice Boltzmann model for simulation of thermal flows on standard lattices, Physical Review E, 85 (2012) 016710.
[48] X.W. Shan, X.F. Yuan, H.D. Chen, Kinetic theory representation of hydrodynamics: a way beyond the Navier-Stokes equation, Journal of Fluid Mechanics, 550 (2006) 413-441.
[49] S.V. Lishchuk, C.M. Care, I. Halliday, Lattice Boltzmann algorithm for surface tension with greatly reduced microcurrents, Physical Review E, 67 (2003) 036701.
[50] I. Halliday, A.P. Hollis, C.M. Care, Lattice Boltzmann algorithm for continuum multicomponent flow, Physical Review E, 76 (2007) 026708.
[51] H. Liu, Y. Zhang, A.J. Valocchi, Modeling and simulation of thermocapillary flows using lattice Boltzmann method, Journal of Computational Physics, 231 (2012) 4433-4453.
[52] Y.W. Kruijt-Stegeman, F.N. van de Vosse, H.E.H. Meijer, Droplet behavior in the presence of

insoluble surfactants, Phys. Fluids, 16 (2004) 2785-2796.

[53] M. Latva-Kokko, D.H. Rothman, Diffusion properties of gradient-based lattice Boltzmann models of immiscible fluids, Physical Review E, 71 (2005) 056702.

[54] A. Gupta, R. Kumar, Effect of geometry on droplet formation in the squeezing regime in a microfluidic T-junction, Microfluidics and Nanofluidics, 8 (2010) 799-812.

[55] J.-J. Xu, H.-K. Zhao, An Eulerian Formulation for Solving Partial Differential Equations Along a Moving Interface, Journal of Scientific Computing, 19 (2003) 573-594.

[56] N.O. Young, J.S. Goldstein, M.J. Block, The motion of bubbles in a vertical temperature gradient, Journal of Fluid Mechanics, 6 (1959) 350-356.

[57] D.P. Ziegler, Boundary conditions for lattice Boltzmann simulations, Journal of Statistical Physics, 71 (1993) 1171-1177.

[58] R. Clift, J.R. Grace, M.E. Weber, Bubbles, Drops, and Particles, (2005).

[59] T. Reis, T.N. Phillips, Lattice Boltzmann model for simulating immiscible two-phase flows, Journal of Physics a-Mathematical and Theoretical, 40 (2007) 4033-4053.